# On the Performance of Turbo Codes in Quasi-Static Fading Channels

M. R. D. Rodrigues, I. Chatzgeorgiou, I. J. Wassell

Laboratory for Communication Engineering
Department of Engineering, University of Cambridge
Cambridge CB3 0FD, United Kingdom
{mrdr3,ic231,ijw24}@cam.ac.uk

R. Carrasco

Communications and Signal Processing Group
School of EE&C Eng., University of Newcastle
Newcastle NE1 7RU, United Kingdom
r.carrasco@ncl.ac.uk

*Abstract*— In this paper, we investigate in detail the performance of turbo codes in quasi-static fading channels both with and without antenna diversity. First, we develop a simple and accurate analytic technique to evaluate the performance of turbo codes in quasi-static fading channels. The proposed analytic technique relates the frame error rate of a turbo code to the iterative decoder convergence threshold, rather than to the turbo code distance spectrum. Subsequently, we compare the performance of various turbo codes in quasi-static fading channels. We show that, in contrast to the situation in the AWGN channel, turbo codes with different interleaver sizes or turbo codes based on RSC codes with different constraint lengths and generator polynomials exhibit identical performance. Moreover, we also compare the performance of turbo codes and convolutional codes in quasi-static fading channels under the condition of identical decoding complexity. In particular, we show that turbo codes do not outperform convolutional codes in quasi-static fading channels with no antenna diversity; and that turbo codes only outperform convolutional codes in quasi-static fading channels with antenna diversity.

## I. Introduction

Berrou *et al.* originally conceived turbo codes over a decade ago [1]. Turbo codes have since been proposed for a variety of wireless applications including mobile and fixed wireless systems, owing to their spectacular performance.

Turbo codes have been shown to be very powerful in the additive white Gaussian noise (AWGN) channel [1]. Turbo codes have also been shown to perform very well in rapidly fading channels [2], but to perform less well in slow fading channels [3]. In rapidly fading channels, coding together with interleaving techniques are used to spread consecutive code bits over multiple independently fading blocks to improve performance. However, in slow fading channels coding together with interleaving techniques cannot in general be used in an effective manner because delay and latency considerations limit the depth of interleaving. This situation compromises in particular the performance of turbo codes because occasional deep fades cause severe error propagation in the iterative decoding process [4].

This paper investigates in detail the performance of turbo codes in quasi-static fading channels both with and without antenna diversity. In the quasi-static fading channel model, the channel response is constant over the length of a data frame, and varies independently from frame to frame. This channel model is thus representative of wireless channels exhibiting extremely slow fading conditions, such as the important broadband fixed wireless access (FWA) channel.

This paper is organized as follows: Section II introduces the system model. Section III develops an analytic technique to evaluate the performance of turbo codes in quasi-static fading channels both with and without antenna diversity. Section IV investigates the performance of turbo codes in quasi-static fading channels both with and without antenna diversity. In particular, we compare the performance of different turbo codes. We also compare the performance of turbo codes and convolutional codes under the condition of identical decoding complexity. Finally, section V summarizes the main contributions of this paper.

## II. System Model

Fig. 1 depicts the communications system model. We consider both single antenna systems ($N_T=N_R=1$), which do not exploit space diversity, as well as multiple antenna systems ($N_T,N_R>1$), which do exploit space diversity.

At the transmitter, the information bits are turbo encoded. The turbo encoder consists of the parallel concatenation of two recursive systematic convolutional (RSC) encoders with rate 1/2, as described in [1]. Alternate puncturing of the parity bits transforms the conventional 1/3 rate code into a 1/2 rate code. The mapper maps groups of two bits into one of four complex symbols from a unit power Gray coded QPSK constellation.

In single transmit antenna systems ($N_T=1$), the space-time processing block does not further process the mapped symbols; instead, the mapped symbols are directly sent to the modulator.

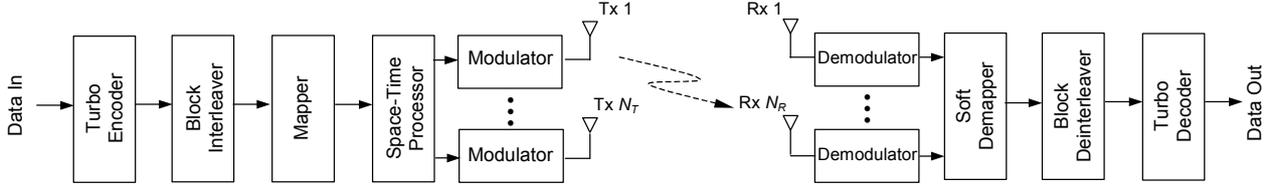

Figure 1. Communications system model.

However, in multiple transmit antenna systems ($N_T$>1), the space-time processing block will further process the mapped symbols. In particular, the space-time processor generates a space-time block code (STBC) according to the generator matrices $\mathbf{G}_2$, $\mathbf{G}_3$ or $\mathbf{G}_4$ given by [5,6]

$$\mathbf{G}_2 = \begin{bmatrix} x_1 & x_2 \\ -x_2^* & x_1^* \end{bmatrix}, \quad (1)$$

$$\mathbf{G}_3 = \begin{bmatrix} x_1 & x_2 & x_3 \\ -x_2 & x_1 & -x_4 \\ -x_3 & x_4 & x_1 \\ -x_4 & -x_3 & x_2 \\ x_1^* & x_2^* & x_3^* \\ -x_2^* & x_1^* & -x_4^* \\ -x_3^* & x_4^* & x_1^* \\ -x_4^* & -x_3^* & x_2^* \end{bmatrix}, \quad (2)$$

$$\mathbf{G}_4 = \begin{bmatrix} x_1 & x_2 & x_3 & x_4 \\ -x_2 & x_1 & -x_4 & x_3 \\ -x_3 & x_4 & x_1 & -x_2 \\ -x_4 & -x_3 & x_2 & x_1 \\ x_1^* & x_2^* & x_3^* & x_4^* \\ -x_2^* & x_1^* & -x_4^* & x_3^* \\ -x_3^* & x_4^* & x_1^* & -x_2^* \\ -x_4^* & -x_3^* & x_2^* & x_1^* \end{bmatrix}, \quad (3)$$

where $x_1$, $x_2$, $x_3$ and $x_4$ denote modulation symbols. The rows of the matrices represent symbols transmitted in different time slots, whereas the columns of the matrices represent symbols transmitted by different antennas. Essentially, a total of $K \times N_T$ symbols obtained from the original $K'$ modulation symbols are transmitted during $K$ time slots by $N_T$ transmit antennas. Note that $\mathbf{G}_2$, $\mathbf{G}_3$ or $\mathbf{G}_4$ are appropriate for two, three and four transmit antennas, respectively, and for an arbitrary number of receive antennas. Note also that $\mathbf{G}_2$ is rate $K'/K=1$, whereas $\mathbf{G}_3$ and $\mathbf{G}_4$ are rate $K'/K=1/2$. Single antenna systems (where $N_T=1$ and $K'=K=1$) are a special case of multiple transmit antenna systems (where $N_T$>1 and $K',K$>1). Thus, in the sequel both single as well as multiple transmit antenna systems are treated under the same framework.

The signal is distorted by a frequency-flat quasi-static fading channel as well as AWGN. Consequently, the relation between the complex receive symbols and the complex transmit symbols associated with a specific STBC frame can be written as follows[1]

$$\mathbf{r} = \mathbf{hs} + \mathbf{n}, \quad (4)$$

where

$$\mathbf{r} = \begin{bmatrix} r_1(1) & r_1(2) & \cdots & r_1(K) \\ r_2(1) & r_2(2) & \cdots & r_2(K) \\ \vdots & \vdots & \ddots & \vdots \\ r_{N_R}(1) & r_{N_R}(2) & \cdots & r_{N_R}(K) \end{bmatrix}, \quad (5)$$

$$\mathbf{s} = \begin{bmatrix} s_1(1) & s_1(2) & \cdots & s_1(K) \\ s_2(1) & s_2(2) & \cdots & s_2(K) \\ \vdots & \vdots & \ddots & \vdots \\ s_{N_T}(1) & s_{N_T}(2) & \cdots & s_{N_T}(K) \end{bmatrix}, \quad (6)$$

$$\mathbf{n} = \begin{bmatrix} n_1(1) & n_1(2) & \cdots & n_1(K) \\ n_2(1) & n_2(2) & \cdots & n_2(K) \\ \vdots & \vdots & \ddots & \vdots \\ n_{N_R}(1) & n_{N_R}(2) & \cdots & n_{N_R}(K) \end{bmatrix}, \quad (7)$$

and

$$\mathbf{h} = \begin{bmatrix} h_{1,1} & h_{1,2} & \cdots & h_{1,N_T} \\ h_{2,1} & h_{2,2} & \cdots & h_{2,N_T} \\ \vdots & \vdots & \ddots & \vdots \\ h_{N_R,1} & h_{N_R,2} & \cdots & h_{N_R,N_T} \end{bmatrix}. \quad (8)$$

Here, $r_j(k)$ denotes the complex receive symbol at time slot $k$ and receive antenna $j$, $s_i(k)$ denotes the complex transmit symbol at time slot $k$ and transmit antenna $i$, $h_{j,i}$ denotes the channel random gain from transmit antenna $i$ to receive antenna $j$ (note that $h_{j,i}$ is independent of time slot $k$), and $n_j(k)$ denotes the noise random variable at time slot $k$ and receive antenna $j$. The channel random gains are uncorrelated circularly symmetric complex Gaussian with zero mean and unit variance; the noise random variables are uncorrelated circularly symmetric complex Gaussian with mean zero and variance $N_T$/SNR, where SNR denotes the signal-to-noise ratio per receive antenna.

---
[1] We focus without loss of generality on the first space-time block code frame.

At the receiver, the soft demapper demaps the complex symbols into soft bits. In particular, the soft demapper computes the log-likelihood ratio (LLR) given by

$$L_D(b_m(k)|\mathbf{r}) = \ln \frac{\text{Prob}\{b_m(k)=1|\mathbf{r}\}}{\text{Prob}\{b_m(k)=0|\mathbf{r}\}}, \quad (9)$$

where $b_m(k)$ is the $m$th bit conveyed by the $k$th modulation symbol. The LLR in (9) is also given by

$$L_D(b_m(k)|\mathbf{r}) = \ln \frac{\sum_{\mathbf{s}\in\mathbf{s}^+} p(\mathbf{r}|\mathbf{s})\text{Prob}\{\mathbf{s}\}}{\sum_{\mathbf{s}\in\mathbf{s}^-} p(\mathbf{r}|\mathbf{s})\text{Prob}\{\mathbf{s}\}} = \ln \frac{\sum_{\mathbf{s}\in\mathbf{s}^+} p(\mathbf{r}|\mathbf{s})\prod_{m'=1}^{2}\prod_{k'=1}^{K'} \text{Prob}\{b_{m'}(k')\}}{\sum_{\mathbf{s}\in\mathbf{s}^-} p(\mathbf{r}|\mathbf{s})\prod_{m'=1}^{2}\prod_{k'=1}^{K'} \text{Prob}\{b_{m'}(k')\}}$$

$$= \underbrace{\ln \frac{\text{Prob}\{b_m(k)=1\}}{\text{Prob}\{b_m(k)=0\}}}_{\textit{a priori}\ \text{information, } L_A(b_m(k))} + \underbrace{\ln \frac{\sum_{\mathbf{s}\in\mathbf{s}^+} p(\mathbf{r}|\mathbf{s})\prod_{m'=1}^{2}\prod_{\substack{k'=1\\m',k'\ne m,k}}^{K'} \text{Prob}\{b_{m'}(k')\}}{\sum_{\mathbf{s}\in\mathbf{s}^-} p(\mathbf{r}|\mathbf{s})\prod_{m'=1}^{2}\prod_{\substack{k'=1\\m',k'\ne m,k}}^{K'} \text{Prob}\{b_{m'}(k')\}}}_{\text{extrinsic information, } L_E(b_m(k)|\mathbf{r})}, \quad (10)$$

where $\mathbf{s}^+$ is the set of matrices of transmit symbols $\mathbf{s}$ such that $b_m(k)=1$ (i.e., $\mathbf{s}^+=\{\mathbf{s}: b_m(k)=1\}$), $\mathbf{s}^-$ is the set of matrices of transmit symbols $\mathbf{s}$ such that $b_m(k)=0$ (i.e., $\mathbf{s}^-=\{\mathbf{s}: b_m(k)=0\}$), and the probability density function $p(\mathbf{r}|\mathbf{s})$ is given by

$$p(\mathbf{r}|\mathbf{s}) = \frac{1}{(2\pi \cdot N_T \cdot \text{SNR}^{-1})^{KN_R}} e^{-\frac{Tr((\mathbf{r}-\mathbf{hs})^H(\mathbf{r}-\mathbf{hs}))}{N_T \cdot \text{SNR}^{-1}}}. \quad (11)$$

Note that the LLR is the sum of the *a priori* information and the extrinsic information, i.e.,

$$L_D(b_m(k)|\mathbf{r}) = L_A(b_m(k)) + L_E(b_m(k)|\mathbf{r}). \quad (12)$$

The *a priori* information is equal to zero, i.e.,

$$L_A(b_m(k)) = 0. \quad (13)$$

The extrinsic information is a function of the STBC scheme. In particular, the extrinsic information expression can be further simplified owing to the orthogonal properties of $\mathbf{G}_2$, $\mathbf{G}_3$ and $\mathbf{G}_4$.

For example, in the single antenna case ($N_T=N_R=1$) with no STBC ($K'=K=1$) it follows that

$$L_E(b_1(1)|r_1(1)) = 4\cdot\text{SNR}\cdot|h_{1,1}|\cdot\text{Re}\{e^{-j\angle h_{1,1}}r_1(1)\}, \quad (14)$$

$$L_E(b_2(1)|r_1(1)) = 4\cdot\text{SNR}\cdot|h_{1,1}|\cdot\text{Im}\{e^{-j\angle h_{1,1}}r_1(1)\}. \quad (15)$$

In the multiple antenna case ($N_T=2, N_R\ge 1$) with the STBC specified by $\mathbf{G}_2$ ($K'=K=2$) it follows that

$$L_E(b_1(1)|\mathbf{r}) = 4\cdot\text{SNR}\cdot\text{Re}\left\{\sum_{n_r=1}^{N_R}(h_{1,n_r})^*\cdot r_{n_r}(1) + h_{2,n_r}\cdot(r_{n_r}(2))^*\right\}, \quad (16)$$

$$L_E(b_2(1)|\mathbf{r}) = 4\cdot\text{SNR}\cdot\text{Im}\left\{\sum_{n_r=1}^{N_R}(h_{1,n_r})^*\cdot r_{n_r}(1) + h_{2,n_r}\cdot(r_{n_r}(2))^*\right\}, \quad (17)$$

$$L_E(b_1(2)|\mathbf{r}) = 4\cdot\text{SNR}\cdot\text{Re}\left\{\sum_{n_r=1}^{N_R}(h_{2,n_r})^*\cdot r_{n_r}(1) - h_{1,n_r}\cdot(r_{n_r}(2))^*\right\}, \quad (18)$$

$$L_E(b_2(2)|\mathbf{r}) = 4\cdot\text{SNR}\cdot\text{Im}\left\{\sum_{n_r=1}^{N_R}(h_{2,n_r})^*\cdot r_{n_r}(1) - h_{1,n_r}\cdot(r_{n_r}(2))^*\right\}. \quad (19)$$

Finally, the soft bits are turbo decoded. The turbo decoder uses the optimal log-domain maximum *a posteriori* (log-MAP) algorithm [7].

## III. SYSTEM ANALYSIS

El Gamal *et al.* have previously devised a simple model to characterize the operation of the turbo iterative decoder [4]. In particular, they have shown that for an energy per bit-to-noise power spectral density ratio $\gamma_b=E_b/N_0$ lower than an iterative decoder convergence threshold $\gamma_{th}=E_{th}/N_0$, the decoder error probability is bounded away from zero independently of the number of decoding iterations. On the other hand, for $\gamma_b$ higher than $\gamma_{th}$, the decoder error probability approaches zero as the number of decoding iterations approach infinity. Here, we exploit this simple model to determine frame error rate expressions for turbo codes in quasi-static fading channels both with and without antenna diversity.

In the single transmit single receive antenna situation errors occur if the instantaneous $\gamma_b$ is less than or equal to $\gamma_{th}$. The channel gain between the transmit and the receive antenna is a complex Gaussian random variable. Consequently, the instantaneous value of $\gamma_b$ is chi-square distributed with two degrees of freedom, i.e.,

$$p(\gamma_b) = \frac{1}{\bar{\gamma}_b}e^{-\frac{\gamma_b}{\bar{\gamma}_b}}, \quad \gamma_b \ge 0, \quad (20)$$

where $\bar{\gamma}_b$ is the average value of $\gamma_b$. Thus, we approximate the frame error rate of the turbo code as

$$P_f = P(\gamma_b \le \gamma_{th}) = \int_0^{\gamma_{th}} p(\gamma_b)d\gamma_b = 1 - e^{-\frac{\gamma_{th}}{\bar{\gamma}_b}}. \quad (21)$$

In the multiple transmit multiple receive antenna situation errors also occur if the instantaneous $\gamma_b$ is less than or equal to $\gamma_{th}$. The channel gains between the various transmit and receive antennas are also complex Gaussian random variables. Consequently, the instantaneous value of $\gamma_b$ is chi-square distributed with $2N_TN_R$ degrees of freedom by virtue of the maximal ratio combining operation associated with the soft demapping operation (e.g., see (16)-(19)), i.e.,

$$p(\gamma_b) = \frac{1}{(N_TN_R-1)!\bar{\gamma}_c^{N_TN_R}}\gamma_b^{N_TN_R-1}e^{-\frac{\gamma_b}{\bar{\gamma}_c}}, \quad \gamma_b \ge 0, \quad (22)$$

where $\bar{\gamma}_c = \bar{\gamma}_b/N_TN_R$ and $\bar{\gamma}_b$ is the average value of $\gamma_b$. Thus, we approximate the frame error rate of a turbo code as

$$P_f = P(\gamma_b \le \gamma_{th}) = \int_0^{\gamma_{th}} p(\gamma_b)d\gamma_b = 1 - e^{-\frac{\gamma_{th}}{\bar{\gamma}_c}}\sum_{k=0}^{N_TN_R-1}\frac{1}{k!}\left(\frac{\gamma_{th}}{\bar{\gamma}_c}\right)^k. \quad (23)$$

The iterative decoder convergence threshold $\gamma_{th}$ can be determined with charts relating the SNR of the extrinsic information [4]. This iterative decoder convergence threshold depends on the structure of the constituent codes (e.g., constituent RSC code rate, constraint length and generator polynomials), rather than that of the composite code.

## IV. RESULTS

This section investigates the performance of turbo codes in quasi-static fading channels both with and without antenna diversity, by analysis and simulation. The turbo encoder uses two identical terminated RSC encoders with rate 1/2, octal generator polynomial (1,5/7) or (1,21/37), and an interleaver size $L$=1024 or 4096. Alternate puncturing of the parity bits transforms the conventional 1/3 rate turbo code into a 1/2 rate turbo code. The turbo decoder uses the log-MAP algorithm with 7 iterations. The iterative decoder convergence threshold for the 1/2 rate turbo code based on RSC codes with generator polynomial (1,5/7) is $\gamma_{th}$=0.77 dB, whereas that for the 1/2 rate turbo code based on RSC codes with generator polynomial (1,21/37) is $\gamma_{th}$=0.57 dB [4].

Figs. 2 and 3 show that, unlike the AWGN channel, turbo codes based on RSCs with different generator polynomials exhibit almost identical performance in quasi-static fading channels for FERs down to $10^{-3}$. This is due to the fact that in this regime turbo code performance in quasi-static fading channels is governed mainly by the convergence characteristics of the iterative decoder, rather than the distance spectrum of the code. Moreover, turbo codes in general, and these two turbo codes in particular, exhibit similar convergence thresholds.

Figs. 4 and 5 also show that, in contrast to the situation in the AWGN channel, turbo codes with different interleaver sizes exhibit identical performance for FERs down to $10^{-3}$. Once again, this is also due to the fact that in this regime turbo code performance in quasi-static fading channels is governed mainly by the convergence characteristics of the iterative decoder, rather than the distance spectrum of the code. Moreover, the interleaver size only affects the distance spectrum of the turbo code, rather than the iterative decoder convergence threshold.

We also observe that analytic results agree very well with simulation results in the various single transmit single receive antenna as well as multiple transmit multiple receive antenna system scenarios. This confirms once again that the performance of turbo codes in quasi-static fading channels depends primarily on the iterative decoder convergence characteristics.

Finally, it is also interesting to compare the performance of turbo codes and convolutional codes in quasi-static fading channels both with and without antenna diversity, under the condition of identical decoding complexity. Here, we consider a 1/2 rate turbo code based on RSCs with generator polynomials (1,5/7), turbo interleaver size $L$=1024 or $L$=4096, and decoded using the log-MAP algorithm with 7 iterations. We also consider a 1/2 rate convolutional code based on an RSC with generator polynomial (1,753/561), and decoded using the Viterbi algorithm. Note that these two configurations exhibit identical decoding complexity in terms of number of equivalent addition operations [8]. Figs. 6 and 7 show that turbo codes do not outperform convolutional codes in quasi-static fading channels with no antenna diversity. Indeed, turbo codes only outperform convolutional codes in quasi-static fading channels with antenna diversity. The figures also show that these results are independent of the turbo interleaver size.

## V. CONCLUSIONS

In this paper, we have investigated in detail the performance of turbo codes in quasi-static fading channels both with and without antenna diversity. First, we have developed a simple and accurate analytic technique to evaluate the performance of turbo codes in quasi-static fading channels. The proposed analytic technique relates the frame error rate of a turbo code to the iterative decoder convergence threshold, rather than the turbo code distance spectrum. Subsequently, we have shown that, in contrast to the situation in the AWGN channel, turbo codes with different interleaver sizes or turbo codes based on RSC codes with different constraint lengths and generator polynomials exhibit identical performance. We have also shown that, under the condition of identical decoding complexity, turbo codes do not outperform convolutional codes in quasi-static fading channels with no antenna diversity; and that turbo codes only outperform convolutional codes in quasi-static fading channels with antenna diversity.

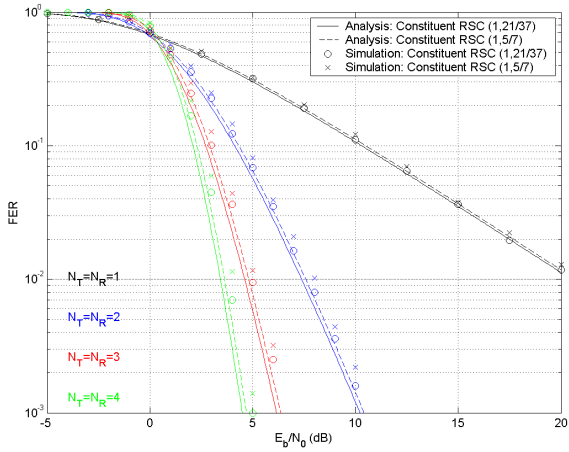

Fig. 2. Frame error rates for turbo codes with different constituent RSC generator polynomials in quasi-static fading channels. Turbo interleaver size $L$=1024.

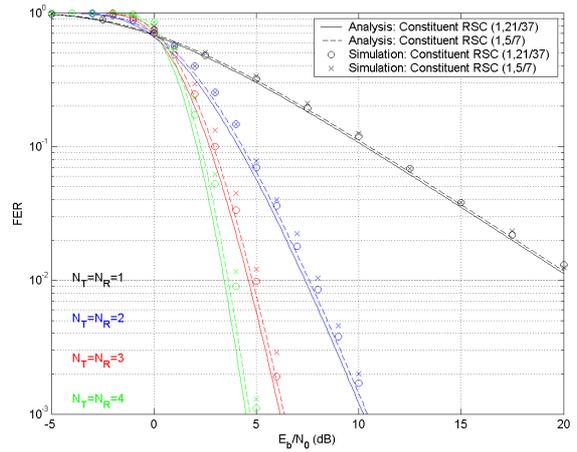

Fig. 3 Frame error rates for turbo codes with different constituent RSC generator polynomials in quasi-static fading channels. Turbo interleaver size $L$=4096.

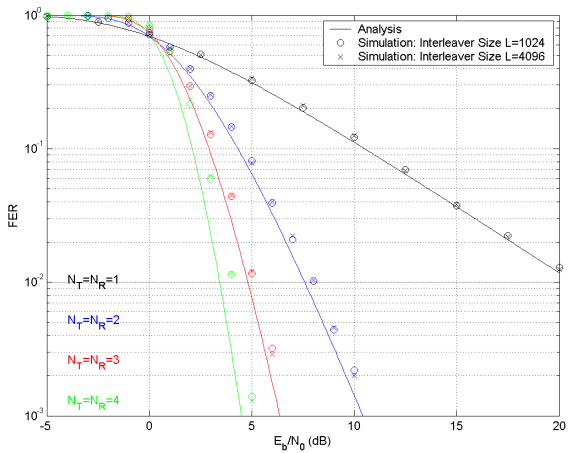

Fig. 4 Frame error rates for turbo codes with different interleaver sizes in quasi-static fading channels. Constituent RSC generator polynomial (1,5/7).

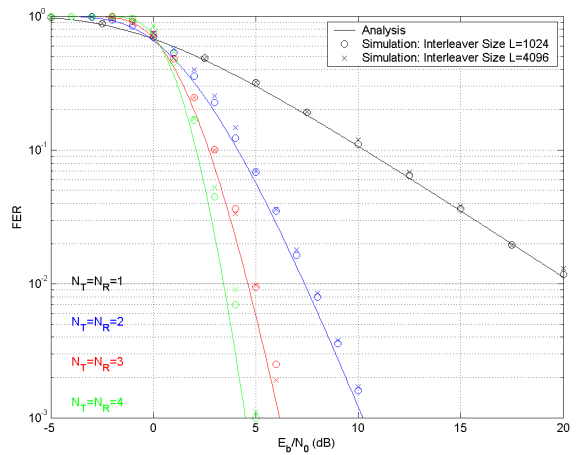

Fig. 5 Frame error rates for turbo codes with different interleaver sizes in quasi-static fading channels. Constituent RSC generator polynomial (1,21/37).

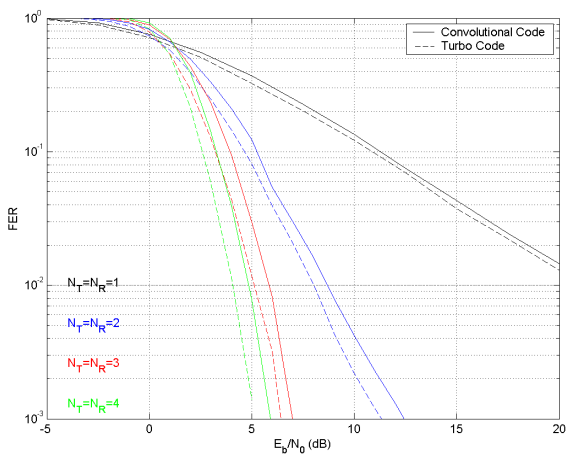

Fig. 6 Simulated frame error rates for turbo codes and convolutional codes in quasi-static fading channels. Turbo interleaver size $L$=1024.

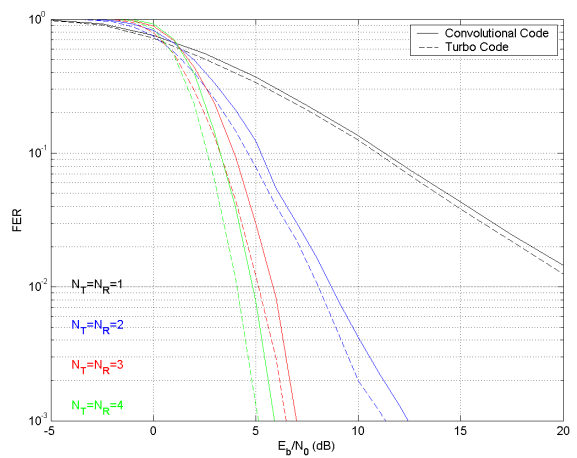

Fig. 7 Simulated frame error rates for turbo codes and convolutional codes in quasi-static fading channels. Turbo interleaver size $L$=4096.